\documentclass[aps,prc,groupedaddress]{revtex4}
\usepackage[pdftex]{color,graphicx}
\begin{document}

\title{The $\phi$ meson at high temperatures and densities}

\author{Gojko Vujanovic\footnote{Current address: Department of Astronomy \& Astrophysics, University of Toronto, 50 St. George Street, Toronto, Ontario M5S 3H4, Canada, $\mathrm{email: vujanovic@astro.utoronto.ca}$}, J\"org Ruppert, and Charles Gale}

\affiliation{Department of Physics, McGill University, 3600 University Street, Montreal, Quebec H3A 2T8, Canada}

\date{\today}

\begin{abstract}
The spectral density of the $\phi$ meson in a hot bath of nucleons and pions is calculated by relating the vector meson self-energy to the forward scattering amplitude (FSA), which is constrained by experimental data. Dispersion techniques are used to verify the relationship between real and the imaginary parts of the in-medium self energy. The position of the spectral peak of the $\phi$ meson is found to be shifted from its vacuum position by only a small amount, but  its width is considerably increased. \end{abstract}


\pacs{}


\maketitle


\section{Introduction}

The collision of relativistic heavy ions remains the only practical way to study hot and dense strongly interacting matter in the laboratory, and a vigorous experimental program at the energy frontier is currently being pursued at RHIC, and another will soon begin at the LHC. The study of in-medium properties of vector mesons is a topic of continuing interest, as their spectral properties can be linked directly to the electromagnetic emissivity of the strongly interacting medium \cite{ftft-book}. As real and virtual photons leave their production site essentially unscathed, they represent penetrating probes that can reveal what conditions existed when they were emitted. Note however that electromagnetic signals are continuously being generated: a reliable interpretation of experimental measurements still hinges on our understanding of the space-time evolution of the source.  Many recent theoretical studies have concentrated on the lower mass vector mesons i.e. $\rho$ and $\omega$ (see \cite{gale-kapusta,RW,eletsky-belkacem-kapusta,martell-ellis,rr,vhr} and references therein), owing partly to the availability of precise heavy ion data \cite{ceres,na60} in this invariant mass range.  In this work, we calculate the in-medium modifications of the properties of the $\phi$ meson. This resonance offers the advantage of a well-defined and narrow distribution, making it a good candidate for precision measurements. On the theory front, some previous  investigations have been pursued in various limits of temperature and density with different techniques. A non-exhaustive list of those tools includes: QCD sum rules \cite{hatlee}, effective hadronic Lagrangians \cite{hg,kww,oram,cabrera-vacas,vhr}, and the Nambu-Jona-Lasinio Lagrangian \cite{bmg}. Estimates of the kinematical broadening of the $\phi$ have also been done \cite{bira,holt-haglin}. Experimentally, RHIC measurements have concentrated up to now on the bulk behavior of the $\phi$ \cite{blyth-ma}, even though some spectral information is emerging \cite{phenix}. In this context, a tantalizing result is that of the KEK-PS E325 collaboration \cite{muto} which claims to have measured the first experimental evidence of in-medium modification of the $\phi$ meson in nuclear matter.    

The goal of this work is not yet to directly compare with experimental measurements of the properties of the $\phi$ in relativistic nuclear collisions. As a first step, the size of expected in-medium modifications should be investigated at finite temperatures and densities, in a simpler equilibrium  setting: we report here on such an investigation. 
The paper is organized in the following way: in Section II we discuss the formalism used to calculate the self-energy, together with the connection to the forward scattering amplitude (FSA). In Section III, the methodology to calculate the FSA is given, and the next section features results. We conclude in section V.  

\section{The in-medium self-energy}

For a $\phi$ meson interacting with hadron $a$ in the medium, the retarded self-energy can be expressed as \cite{jeon-ellis,eletsky-ioffe-kapusta}
\begin{equation}
\Pi_{\phi a} \left( p \right)= -\frac{m_\phi m_a T}{\pi p} \int^{\infty}_{m_a}d\omega\ln\left[\frac{1-\exp\left(-\omega_{+}/T\right)}{1-\exp \left(-\omega_{-}/T\right)}\right] f_{\phi a}\left(\frac{m_\phi}{m_a}\omega\right) 
\end{equation}
to leading order in the density of scatterers, in the rest frame of $a$, and on the mass shell of the $\phi$ meson. In the above, $p=\left|{\bf p}\right|$ is the momentum of the $\phi$ meson, $\omega^2=m^2_a+k^2$, $f_{\phi a}$
is the amplitude for the forward scattering of $\phi$ with the field of type $a$, and $T$ is the temperature. Also, $\omega_\pm = \frac{E\omega \pm pk}{m_\phi}$, with $a$ a boson. If $a$ is a fermion with a chemical potential $\mu$, the argument of the logarithm becomes $ \left( 1+\exp\left(-\frac{\left( \omega_{-} - \mu \right)}{T} \right) \right)/ \left( 1+\exp \left( -\frac{ \left( \omega_{+}-\mu \right)}{T} \right) \right) $ \cite{eletsky-belkacem-kapusta}. 

\begin{figure}[!ht]
\begin{center}
\includegraphics[scale=0.7]{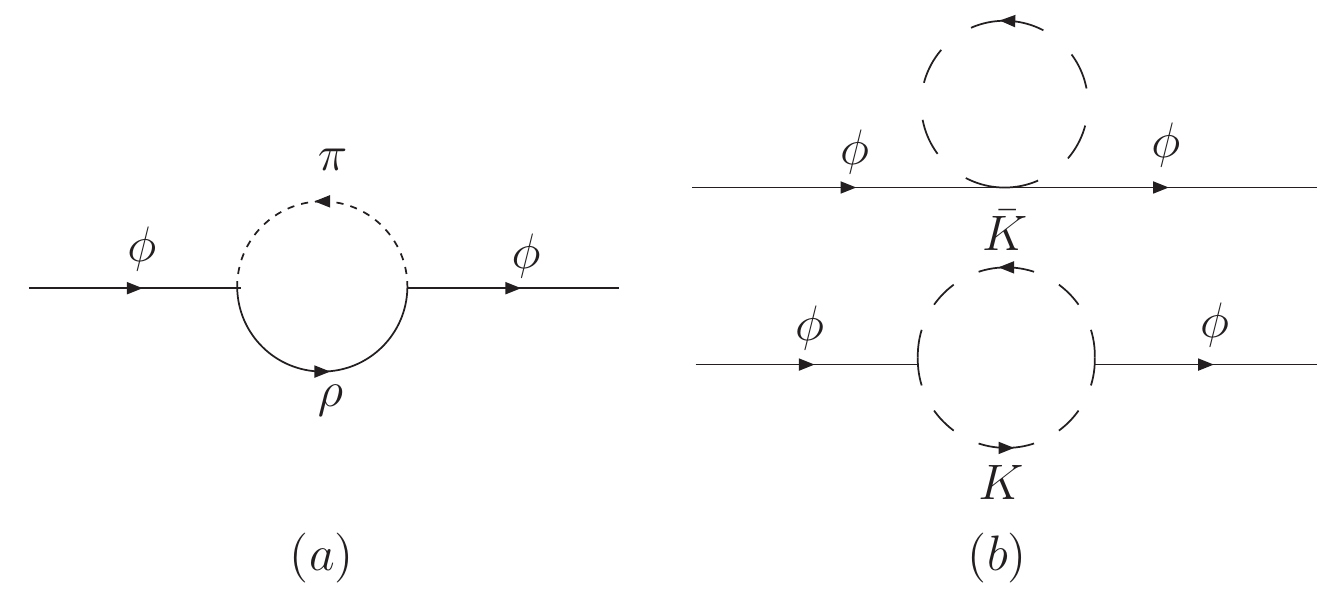}
\end{center}
\caption{The one-loop vacuum self-energy of the $\phi$ meson.}\label{pic:phi_self}
\end{figure}

Turning first to the zero temperature part of the self-energy, we identify two contributions. The first is calculated through the interaction of the $\phi$ meson with a $\rho$ and $\pi$, and is illustrated in Fig. \ref{pic:phi_self} (a).
This interaction is well described by the Wess-Zumino Lagrangian term, with $\omega$-$\phi$ mixing \cite{wess-zumino}:
\begin{equation}
\mathcal{L}_{(\omega,\phi)\rho \pi} = g \epsilon^{\alpha \beta \mu \nu} \partial_\alpha {\bf \rho}_\beta \cdot {\bf \pi} \left( \frac{\partial_\mu \omega^{8}_{\nu}+\sqrt{2}\partial_\mu  \omega^s_{\nu}}{\sqrt{3}} \right) 
\end{equation}
where 
\begin{eqnarray}
\omega^8 & = & \phi \cos\left( \theta_V \right) + \omega \sin \left( \theta_V \right)  \\
\omega^s & = & - \phi \sin\left( \theta_V \right) + \omega \cos \left( \theta_V \right)  
\end{eqnarray}
Using the Wess-Zumino interaction, we calculate the vacuum self-energy of the $\phi$ meson:
\begin{eqnarray}
\Pi^{\rm vac}_{\phi\rightarrow \rho \pi}\left( M \right)  & = & M^2 \frac{g^2}{\left(4\pi\right)^2}\left( \sqrt{\frac{1}{3}} \cos\left(\theta_V \right) - \sqrt{\frac{2}{3}} \sin \left(\theta_V \right)\right)^2 \times {} \nonumber \\ 
& & {} \times  \int^1_0 dx \Delta \left[ -\ln\left( \Delta \right) + \ln \left( 4 \pi \right) + 1 - \gamma_{\rm E} \right] + C \label{eq:self-rho-pi}
\end{eqnarray}
Using the experimentally measured branching ratio, the coupling constant in a given isospin channel is determined to be $g^2=1.766 \times 10^{-4}$ MeV, while the mixing angle is $\theta_V=40.1^{\circ}$ \cite{wess-zumino,haglin-gale}. Note that $\Delta = m^2_\rho - x \left( m^2_\rho - m^2_\pi\right) -x\left(1-x \right)M^2$, $M$ is the invariant mass of the $\phi$ meson, and $\gamma_{\rm E}$ is the Euler-Mascheroni constant. The renormalization constant $C$ is chosen such that ${\rm Re} \left[ \Pi \left( M^2 = m_\phi^2 \right) \right]=0$. 

The second contribution to the vacuum self-energy is given in Fig. \ref{pic:phi_self} (b). Such an interaction is described by the Lagrangian of the form \cite{gale-kapusta,haglin-gale}: 
\begin{equation}
\mathcal{L}_{\phi \rightarrow K \bar{K}} = { \frac{1}{2}}\left|D_\mu K\right|^2 - { \frac{1}{2}}m^2_{K}\left|K\right|^2 -\frac{1}{4} F^{\mu \nu}_{\phi} F^{\phi}_{\mu \nu}+ \frac{1}{2} m^2_{\phi} \phi^{\mu} \phi_{\mu} 
\end{equation}
where $K$ in the complex charged kaon field, $F^{\phi}_{\mu \nu}$ is the $\phi$ field strength and $D_\mu = \partial_\mu - i g_{\phi\rightarrow K\bar{K}} \phi_\mu$ is the covariant derivative. The vacuum self-energy for the $\phi \rightarrow K \bar{K}$ interaction is:
\begin{equation}
\Pi^{\rm vac}_{\phi \rightarrow K \bar{K}} \left( M \right) = \frac{M^2}{3}\frac{g^2_{\phi\rightarrow K \bar{K}}}{\left(4\pi\right)^2}\left\{ \left(1-\frac{4 m^2_K}{M^2} \right)^{3/2} \left[ \ln \left| \frac{1+\sqrt{1-\frac{4 m^2_K}{M^2}}}{1-\sqrt{1-\frac{4 m^2_K}{M^2}}}\right| -i\pi\Theta\left( M^2 - 4m^2_K\right)\right] + \frac{8 m^2_K}{M^2} + \bar{C}\right\} \label{eq:self-kk}
\end{equation}
where the kaon mass is either $m_K=0.4937$ GeV or $m_K=0.4976$ GeV for the charged and neutral Kaons respectively. The values of the coupling are $\frac{g^2_{\phi\rightarrow K \bar{K}}}{4\pi}=1.602$ and 1.682 for the charged and neutral kaons \cite{haglin-gale}. Finally, $\bar{C}$ is determined the same way as $C$ in Eq. (\ref{eq:self-rho-pi}). 

The net self-energy is given by summing over all target species and including the vacuum contributions:
\begin{equation}
\Pi^{\rm net}_\phi (E, p)= \Pi^{\rm vac}_\phi \left( M \right) +  \Pi_{\phi \pi} \left( p \right) +  \Pi_{\phi N}\left( p \right).
\end{equation}
The individual contributions to the self-energy contain the appropriate spin/isospin symmetry factors. Note that the functional dependence of the self-energy in the vacuum case is different than that at finite temperature: the vacuum part of $\Pi$ can solely depend on the invariant mass, $M=\sqrt{E^2-p^2}$. The matter parts do not have such a restriction and in general depend on both $E$ and $p$. However, in this study, the scattering amplitudes are evaluated on the mass shell of the $\phi$ meson. Therefore, the vacuum contribution is evaluated at $M=m_\phi$ and the matter self-energies are only $p$-dependent.

\section{Modeling the forward scattering amplitude}
\subsection{The low energy resonance contribution to the FSA}\label{sec:fsa_low}

We only consider interactions with pions ($\pi$) and nucleons ($N$) as scatterings of the $\phi$ meson contributing to its self-energy. To describe $\phi$'s self-energy, a two-component approach \cite{harari} is used (see also \cite{collins}): while ordinary Reggeons are dual to s-channel resonances, the Pomeron is dual to the background under the resonances. The low energy FSA is therefore composed of two parts, namely a resonance contribution and a background Pomeron term. In the center of mass (cm) frame the low energy FSA reads \cite{eletsky-belkacem-kapusta}
\begin{equation}
f^{\rm cm}_{\phi a}(s) = \frac{1}{2q_{\rm cm}}\sum_{R} W^{R}_{\phi a}\frac{\Gamma_{R\rightarrow \phi a}}{M_{R} - \sqrt{s} - \frac{1}{2}i\Gamma_{R}} - \frac{q_{\rm cm}}{4 \pi s} \frac{1+\exp(-i \pi \alpha_P)}{\sin(\pi \alpha_P)} r_{\phi a}^{P} s^{\alpha_P}. \label{eq:f_low}
\end{equation}
Here the sum ranges over resonances that decay into the $\phi$ meson and the particle $a$, which is either a nucleon or a pion. The mass of the resonance $R$ is $M_R$ and its total width is $\Gamma_R$. As usual,  $s$ is the  Mandelstam variable, and the magnitude of the center of mass momentum can be re-expressed in terms of masses and $s$. Spin/isospin statistics are taken care of by using the averaging factor $W^{R}_{\phi a} = \frac{(2s_R + 1)}{(2s_{\phi}+1)(2s_a+1)} \frac{(2t_R + 1)}{(2t_{\phi} + 1)(2t_a + 1)}$, with $s_i$ being the spin of particle $i$, and $t_i$, the isospin. This averaging procedure will wash out the distinction between longitudinal and transverse spin directions, and also the difference between different charge states.   

The  partial width for a resonance to decay into the $\phi a$ channel is denoted by $\Gamma_{R\rightarrow \phi a}$ in Eq.(\ref{eq:f_low}). In the case were $a$ is a nucleon, the effective width cannot be obtained directly from experiment \cite{pdg}, as no resonances that decay directly into $\phi$ and $N$ have yet been observed \cite{footnote}.  A scaling law first proposed by Lipkin \cite{lipkin} in order to infer the effective width of $\Gamma_{R\rightarrow \phi a}$ is used. Lipkin found that the ratio $R_{\phi/\omega}=\frac{g^2_{\phi \rho \pi}}{g^2_{\omega \rho \pi}}=\frac{g^2_{\phi N N}}{g^2_{\omega N N}}=\frac{\sigma \left( \pi N \rightarrow \phi X \right)}{\sigma \left( \pi N \rightarrow \omega X  \right)}=\frac{ \sigma \left( N N \rightarrow \phi X \right) }{\sigma \left( N N \rightarrow \omega X \right)}$ is a constant that quantifies the experimental deviation from the ideal SU(3) octet-singlet mixing. We will be using the average value of $<R_{\phi/\omega}>$ (averaged over the range $0<\sqrt{s}-m_{V}-m_{N}<10$ GeV), henceforth denoted $R_{\phi/\omega}$. Sibirtsev \textit{et al.} \cite{sibirtsev} have recently obtained $R_{\phi/\omega}= \left( 13.4 \pm 3.2 \right) \times 10^{-3}$ via analysis of $\phi$ and $\omega$ production in $\pi N$ and $N N$ reactions. It is assumed that resonances that can connect to an $\omega N$ state \cite{martell-ellis} will also have a $\phi N$ interaction, with the appropriate kinematic adjustments. However, those resonances will have a reduced branching ratio $B$ by $R_{\phi/\omega}$, i.e. $B_{R \rightarrow \phi N} = R_{\phi/\omega} B_{R \rightarrow \omega N}$. Similarly to what was previously done \cite{eletsky-belkacem-kapusta,martell-ellis}, $\Gamma_{R\rightarrow \phi N}$ is expressed as:  
\begin{eqnarray}
\Gamma_{R \rightarrow \phi N} = \left\{ \begin{array}{ll}
\Gamma_R B_{R \rightarrow \omega N} R_{\phi/\omega} \left( \frac{q_{\rm cm}}{q^R_{\rm cm}} \right)^{2l+1} & \textrm{$q_{\rm cm}$ $\leq$ $q^{R}_{\rm cm}$}\\
\Gamma_R B_{R \rightarrow \omega N} R_{\phi/\omega} & \textrm{$q_{\rm cm}$ $\geq$ $q^{R}_{\rm cm}$} \\
\end{array} \right.
\end{eqnarray}
where $\Gamma_{R}$ is the total width of the resonance $R$, $B_{R \rightarrow \omega N}$ is the branching ratio of the decay $R \rightarrow \omega N$, we take the average value $R_{\phi/\omega} = 13.4\times 10^{-3}$, $q^{R}_{\rm cm} =  \frac{1}{2} \frac{ \sqrt{\left[ M^{2}_{R} - \left( m_\phi + m_a \right)^2 \right] \left[ M^{2}_{R} - \left( m_\phi - m_a \right)^2 \right]} }{ M_{R} }$, and $l$ is the smallest relative angular momentum between $\phi$ and $N$. 

Reactions proceeding via the strong interaction conserve parity; we use this fact to determine $l$. The parity conservation can be stated as 
\begin{equation}
P_{R}=(-1)^{l}P_{\phi}P_{N}
\end{equation}
where $P_R$ is the parity of the resonance, $P_{\phi}$ is the parity of the $\phi$ meson, $P_N$ is the parity of the nucleon, and $(-1)^l$ is the parity of the wavefunction describing the relative angular momentum of $\phi$ and $N$. Parity conservation determines only whether $l$ is even or odd. To determine minimum $l$, we need to use spin and angular momentum addition rule. This rule looks as follows:
\begin{equation}
\hat{j}_R= \hat{l} + \hat{s}_{\phi} + \hat{s}_N  
\end{equation} 
where $\hat{j}_R$ is the total spin operator of the resonance $R$, $\hat{l}$ is the angular momentum operator, $\hat{s}_{\phi}$ is the spin operator of the $\phi$ meson, and $\hat{s}_{N}$ is the spin of the nucleon. The non-vanishing Clebsch-Gordan coefficients will determine the possible values $l$ can take in the even/odd number region delimited by parity conservation equation. We keep only the smallest $l$ that satisfies both spin and parity conditions. The above approach to determine spin and parity is general and also applicable for interactions with pions (instead of nucleons).  

\begin{figure}[!ht]
\begin{center}
\includegraphics[scale=0.7]{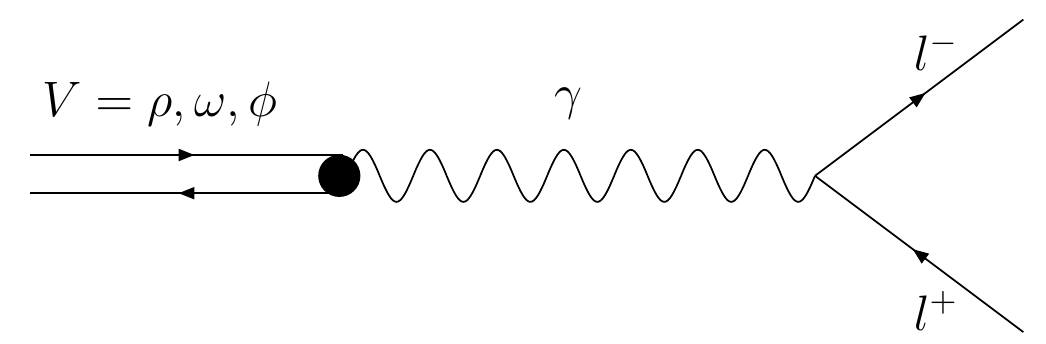}
\end{center}
\caption{Feynman Diagram used in the calculation of the width of a vector meson going into a dilepton.}\label{pic:em_decay}
\end{figure}

Besides the above-threshold resonances, it is also necessary to include subthreshold resonances for they also make a contribution to the FSA. Their contribution however is relatively small. Indeed, including subthreshold resonances changes the imaginary part of the FSA by at most 2\% in the $0<E_{\phi}-m_{\phi}<0.25$ GeV energy range. The change to the real part of the FSA due to subthreshold effects is significantly larger reaching 50\% in the same energy range. Hence it is important to include the subthreshold resonances in the FSA. Note that the overall magnitude of the real part of the FSA in that region is quite small  (see Fig. \ref{pic:re_im_f}), so the aforementioned 50\% change is not as dramatic as  might first appear. To estimate their widths we assume that the vector meson dominance model (VMD) is valid. As in \cite{eletsky-belkacem-kapusta,martell-ellis}, we write the partial width of the subthreshold resonances as $\Gamma_{R \rightarrow \phi N} = q_{\rm cm} \gamma_{R\rightarrow \phi N}$ and $\Gamma_{R\rightarrow \gamma N} = k_{\rm cm} \gamma_{R \rightarrow \gamma N}$, where $k_{\rm cm}$ is the $\gamma N$ center of mass momentum. VMD allows us to relate $\gamma_{R\rightarrow V N}$ (with $V=\rho, \omega, \phi$) to $\gamma_{R\rightarrow \gamma N}$ as follows \cite{ftft-book}:
\begin{equation}
\gamma_{R\rightarrow\gamma N} = 4 \pi \alpha \left( \frac{\gamma_{R\rightarrow\rho N}}{g^2_{\rho}} + \frac{\gamma_{R\rightarrow\omega N}}{g^2_{\omega}}+ \frac{\gamma_{R\rightarrow\phi N}}{g^2_{\phi}}\right). \label{subth_gamma}
\end{equation}
One may evaluate the value of the couplings $g^2_{\rho, \omega, \phi}$ by first writing down the VMD Lagrangian \cite{ftft-book} which governs the  electromagnetic decay of vector mesons:
\begin{equation}
\mathcal{L} = \mathcal{L}_{QED} - \sum_{V = \rho, \omega, \phi} \frac{e}{g_V}m^2_{V}V^{\mu}A_{\mu} - \sum_{V = \rho, \omega, \phi} \frac{1}{4} F^{\mu \nu}_V F^V_{\mu \nu}
\end{equation}
where $\mathcal{L}_{QED}= \bar{\psi}_{l} \left( i\not \! \partial - m_{l}\right) \psi_{l} - \frac{1}{4}F^{\mu \nu}F_{\mu \nu} - \sqrt{4\pi\alpha}\bar{\psi}_{l}\gamma^{\mu}\psi_l A_{\mu}$. The fermionic field for lepton $l$ is  $\psi_l$, $A_{\mu}$ is the photon field with the corresponding field strength tensor $F_{\mu \nu}=\partial_{\mu}A_{\nu}-\partial_{\nu}A_{\mu}$,  $V^{\mu}$ are the different vector meson fields, namely $\rho$, $\omega$ and $\phi$ fields, with the field strength tensor written in an analogous fashion to the photon case and $\alpha$ is the electromagnetic fine-structure constant.  Finally, the different $m_i$ (with $i$ generic) are the masses of the different particles involved. The coupling constants are then easily determined by evaluating the decay width into a lepton pair
\begin{equation}
\Gamma_{V \rightarrow l^{+} l^{-}} = \frac{\alpha^2}{3} \frac{m_{V}}{g^2_V/(4\pi)}\left( 1 + \frac{2m^2_{l}}{m^2_{V}} \right) \left(1 - \frac{4m^2_{l}}{m^2_{V}} \right)^{1/2}.
\end{equation}
which stems from the Feynman amplitude of Fig. \ref{pic:em_decay}.  
The values obtained are $\frac{g^2_{\rho}}{4\pi}=2.54$, $\frac{g^2_{\omega}}{4\pi}=20.5$, and $\frac{g^2_{\phi}}{4\pi}=11.7$ \cite{ioffe}. Inspired by the quark model it is estimated that $\gamma_{R\rightarrow \rho N}\approx \gamma_{R\rightarrow \omega N}$ \cite{ftft-book}. Further, using the scaling law proposed by Lipkin, the relation $\gamma_{R\rightarrow \phi N}\approx R_{\phi/\omega}\gamma_{R\rightarrow\omega N}$ is invoked. These statements allow to simplify Eq. (\ref{subth_gamma}):
\begin{equation}
\gamma_{R\rightarrow \phi N}=\frac{\gamma_{R\rightarrow \gamma N}}{4 \pi \alpha } \left( \frac{1}{g^2_{\rho} R_{\phi/\omega}} + \frac{1}{g^2_{\omega}R_{\phi/\omega}}+ \frac{1}{g^2_{\phi}}\right)^{-1}
\end{equation}
where $\gamma_{\gamma N} = \frac{1}{2} \left( \gamma_{\gamma p} + \gamma_{\gamma n}\right)$, with $\gamma_{\gamma (p,n)}$ being the experimental values quoted in Ref. \cite{pdg}. 
A summary of all the resonances that decay into $\phi N$ is presented in Table \ref{table:phi_N}. 
\begin{table}[!ht]
\caption{Baryon resonances, both above and below threshold, included in the calculation of the $\phi N$ FSA.} 
\centering 
\begin{tabular}{c c c c} 
\hline \hline 
Resonance &  Mass  & Width & Branching ratio \\
          &  (GeV) & (GeV) & ($\times 13.4\cdot10^{-3}$)          \\ [0.5ex] 
\hline 
N(2190) & 2.127 & 0.547 & 0.490 \\
N(2000) & 1.981 & 0.361 & 0.022 \\
N(1900) & 1.900 & 0.498 & 0.390 \\
N(1720) & 1.720 & 0.200 & 1.2$\times 10^{-3}$  \\ 
N(1710) & 1.710 & 0.100 & 0.18$\times 10^{-3}$ \\ 
N(1700) & 1.700 & 0.100 & 0.50$\times 10^{-3}$ \\
N(1680) & 1.685 & 0.130 & 1.5$\times 10^{-3}$  \\
N(1675) & 1.675 & 0.150 & 0.42$\times 10^{-3}$ \\
N(1650) & 1.655 & 0.165 & 0.98$\times 10^{-3}$ \\
N(1535) & 1.535 & 0.150 & 1.9$\times 10^{-3}$  \\
N(1520) & 1.520 & 0.115 & 4.6$\times 10^{-3}$ \\
N(1440) & 1.440 & 0.300 & 0.31$\times 10^{-3}$ \\ 
[1ex] 
\hline 
\end{tabular}
\label{table:phi_N} 
\end{table}
The data in Table \ref{table:phi_N} is mostly from \cite{pdg} except for the resonances N(2000) and N(2190), which are taken from \cite{martell-ellis,spin-5/2}. 

In cases involving the pion, since there are particles quoted by the Particle Data Group that decay into $\phi \pi$, the effective width to be used in Eq. (\ref{eq:f_low}) takes the following form:
\begin{eqnarray}
\Gamma_{R \rightarrow \phi \pi} = \left\{ \begin{array}{ll}
\Gamma_R B_{R \rightarrow \phi \pi} \left( \frac{q_{\rm cm}}{q^R_{\rm cm}} \right)^{2l+1} & \textrm{$q_{\rm cm}$ $\leq$ $q^{R}_{\rm cm}$}\\
\Gamma_R B_{R \rightarrow \phi \pi} & \textrm{$q_{\rm cm}$ $\geq$ $q^{R}_{\rm cm}$} \\
\end{array} \right.
\end{eqnarray}
Here, all the variables are defined analogously to the $\phi N$ case. One can use the above formula directly for all $l\neq 0$.   
For $l=0$ however, Adler's theorem should be fulfilled. We follow here the procedure discussed in \cite{eletsky-belkacem-kapusta}. According to this theorem, the pion scattering amplitude on any hadronic target vanishes when $q_{\rm cm} \rightarrow 0$ in the limit of massless pions. In order to satisfy this theorem, in an effective Lagrangian approach, the coupling of the pion field with other particles is a derivative coupling, $\partial_{\mu} \pi$. Therefore, the matrix element $\mathcal{M}$ goes as $\mathcal{M} \sim g \left( k^{\mu} \right)^{l}$. Letting $l=0$, i.e. Breit-Wigner contribution for s-waves, we notice that the effective width $\Gamma_{R \rightarrow \phi \pi}$ doesn't vanish as $q_{\rm cm} \rightarrow 0$, which it should in order to obey Adler's theorem. Consequently, the effective width for $l=0$ now must be rewritten as in \cite{eletsky-belkacem-kapusta}:  
\begin{eqnarray}
\Gamma_{R \rightarrow \phi \pi} = \left\{ \begin{array}{ll}
\Gamma_R B_{R \rightarrow \phi \pi} \left( \frac{s-m_{\phi}^2-m_{\pi}^2}{s_0-m_{\phi}^2-m_{\pi}^2} \right)^2 & \textrm{$s$ $\leq$ $s_0$}\\
\Gamma_R B_{R \rightarrow \phi \pi} & \textrm{$s$ $\geq$ $s_0$} \\
\end{array} \right.
\end{eqnarray}
with $s_0=\left( m_{\phi} + m_{\pi} + m_{\rho} \right)^2$. The established resonance that decays into $\phi \pi$ is $\rho(1450)$ having a mass of 1.480 GeV, a width of 0.4 GeV, and a branching ratio of 32.5\% \cite{pdg}. We have not included other resonances since their branching ratios are both experimentally uncertain and numerically small. 

We have gone through all of the details in Eq. (\ref{eq:f_low}) except for the second term, the Pomeron background contribution. This is  discussed in section \ref{sec:fsa_high}: it functional behavior resembles that of the high energy FSA. 

\subsection{The FSA at high energies: a Regge parametrisation} \label{sec:fsa_high}

The high energy FSA's can be well described by a Regge parametrisation of the form:
\begin{equation}
f^{\rm cm}_{\phi a} \left(s\right)= -\frac{q_{\rm cm}}{4 \pi s} \sum_{i} \left[ \frac{1+\exp(-i \pi \alpha_i)}{\sin(\pi \alpha_i)} \right] r_{\phi a}^{i} s^{\alpha_i}. \label{eq:f_high}
\end{equation}
where, in our application, $a$ is a nucleon or a pion. We consider only two terms to be present in the sum in Eq. (\ref{eq:f_high}): a Pomeron term $P$ and a Regge term $P^{'}$. Such a construction of the high energy FSA is motivated by previous work done by Donnachie and Landshoff \cite{Donnachie-Landshoff}, where it is shown that such a Regge parametrization seems to well describe cross section data. Using the optical theorem, $\sigma_{\phi a} \left(s\right)= \frac{4\pi}{q_{\rm cm}} {\rm Im }\left[f^{\rm cm}_{\phi a}\left( s \right)\right]$, we see that the parametrization in Eq.(\ref{eq:f_high}) reduces to the form used by Donnachie and Landshoff. At high energies, scattering is dominated by contributions from individual quarks, hence the additive quark model is applicable \cite{eletsky-belkacem-kapusta}. Therefore, averaging over charged states, we take the cross section $\sigma_{\phi N} \simeq \sigma_{\pi N}$. The residues $r_i$ and the intercepts $\alpha_i$ of the $i$th Regge trajectory are $\alpha_P=1.093$ and $\alpha_{P^{'}}=0.642$ with $r^{\phi N}_{P}=11.88$ and $r^{\phi N}_{P^{'}}=28.59$ \cite{eletsky-belkacem-kapusta,martell-ellis}. Also, we approximate $\sigma_{\phi \pi} \simeq \sigma_{\pi \pi}$, averaged over charged states of course. For the $\phi \pi$ decays, the residues are $r^{\phi \pi}_{P}=7.508$ and $r^{\phi \pi}_{P^{'}}=12.74$ \cite{eletsky-belkacem-kapusta,martell-ellis}. The intercepts $\alpha_i$ however are universal quantities. The parameters for the Pomeron obtained here are also used for the Pomeron background term in Eq.(\ref{eq:f_low}).  One {\it a posteriori} verification of our procedure is the test provided by dispersion techniques in the next section.

\section{Results}

The properties of the $\phi$ meson in the rest frame of the heat bath will be calculated. The FSA where the particle $a$ is at rest is related to the center of mass FSA by $f_{\phi a} \left( E_\phi \right)=  \frac{\sqrt{s}}{m_a}f^{\rm cm}_{\phi a} (s)$ where $E_\phi - m_\phi = \frac{s-\left( m_\phi+m_a \right)^2}{2 m_a}$. In order to have a complete description of the FSA, its low and high energy parts will be matched smoothly. There is no strict theoretical guidance that helps to perform this matching in a unique way. The interpolating procedure chosen is defined presently: the matching is done via one half-sided function $g\left(E_\phi\right)$ for both the real and imaginary parts:
\begin{eqnarray}
g\left(E_\phi\right) = \left\{ \begin{array}{ll}
\exp\left[ \frac{\left( E_\phi - b \right)}{\sigma}\right] & \textrm{ $E_\phi$ $\geq$ $b$ } \\
1 & \textrm{$E_\phi$ $\leq$ $b$}\\
\end{array} \right.
\end{eqnarray}
where $b$ and $\sigma$ are free parameters. The matched function takes the form
\begin{equation}
f^{total}_{\phi a} \left(E_\phi \right) = g\left( E_\phi \right) f^{low}_{\phi a} \left( E_\phi \right) +  \left(1-g\left(E_\phi \right)\right) f^{high}_{\phi a} \left( E_\phi \right)
\end{equation}
One may verify how well the matching is done by using a dispersion integral formula relating the real part of the total FSA to a principal value integral over its imaginary part \cite{eletsky-belkacem-kapusta}:
\begin{equation}
{\rm Re}\left[f_{\phi a} \left(E_\phi \right) \right] = {\rm Re} \left[f_{\phi a} \left( 0 \right) \right] + \frac{2E^{2}_{\phi}}{\pi} \mathrm{P.V.} \int^{\infty}_{m_\phi} \frac{{\rm Im} \left[ f_{\phi a}\left( E^{\prime}\right) \right]  d E^{\prime}}{E^{\prime}\left( E^{\prime}+E_{\phi}\right) \left( E^{\prime} - E_{\phi}\right)}. 
\end{equation}
The free parameters $b$ and $\sigma$ are chosen such that ${\rm Re} \left[ f_{\phi a} \right]$ minus the dispersion integral is as close to a constant as possible. 

\begin{figure}[!ht]
\begin{center}
\includegraphics[scale=0.45]{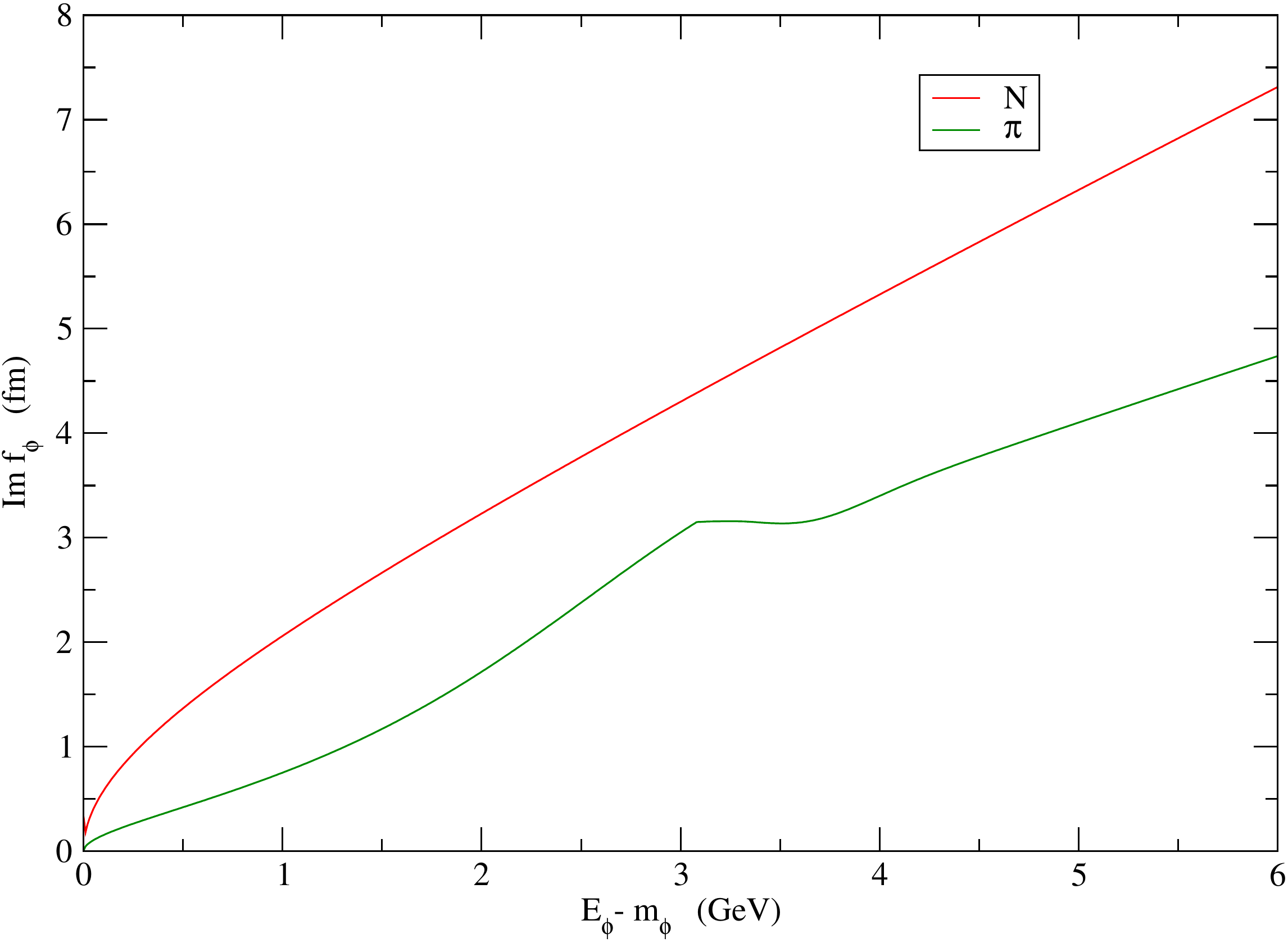}
\includegraphics[scale=0.45]{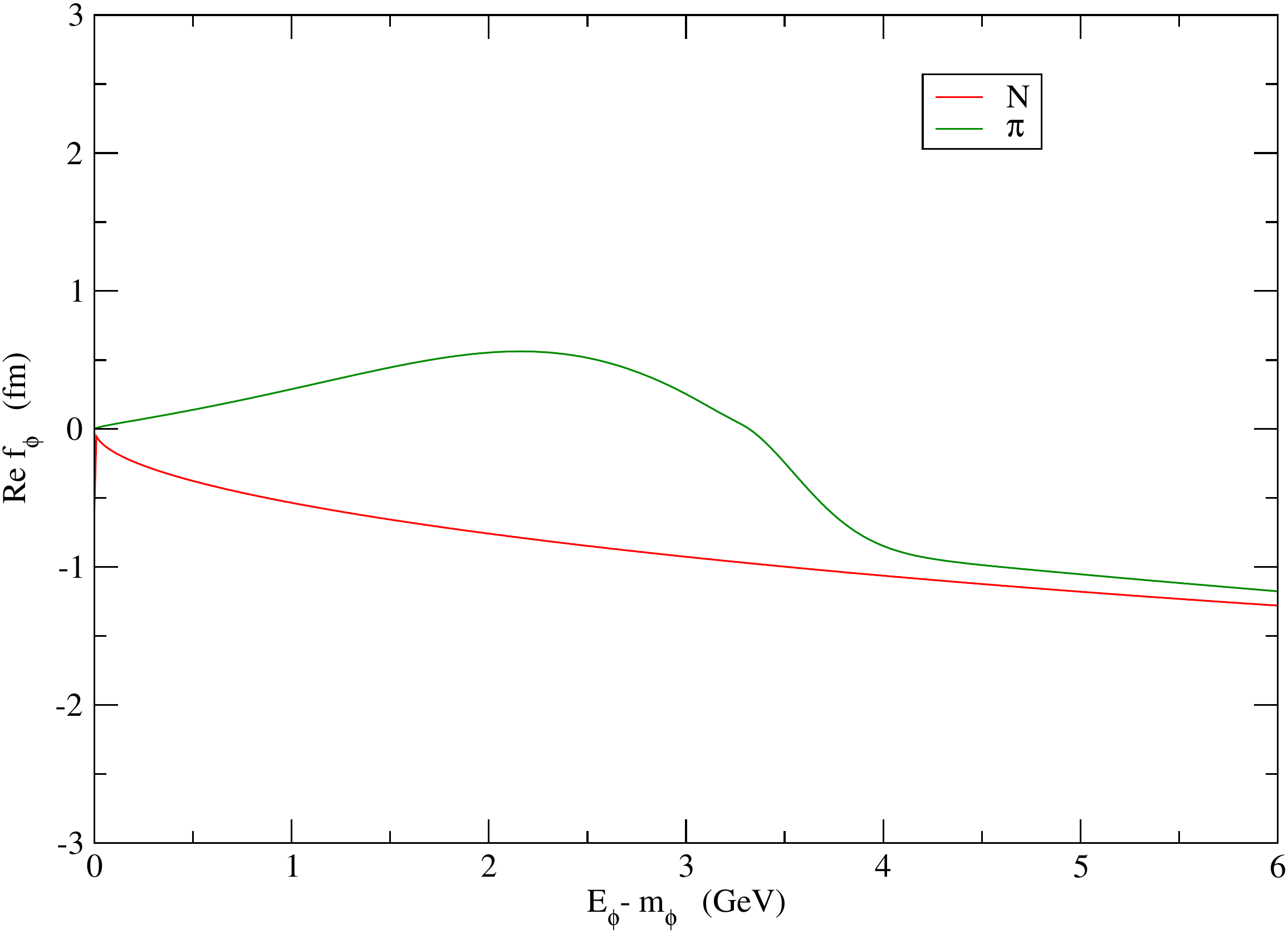}
\end{center}
\caption{(a) The imaginary and (b) the real part of the FSA for both $\phi N$ (red line) scattering and $\phi \pi$ scattering (green line).}\label{pic:re_im_f}
\end{figure}

\begin{figure}[!ht]
\begin{center}
\includegraphics[scale=0.45]{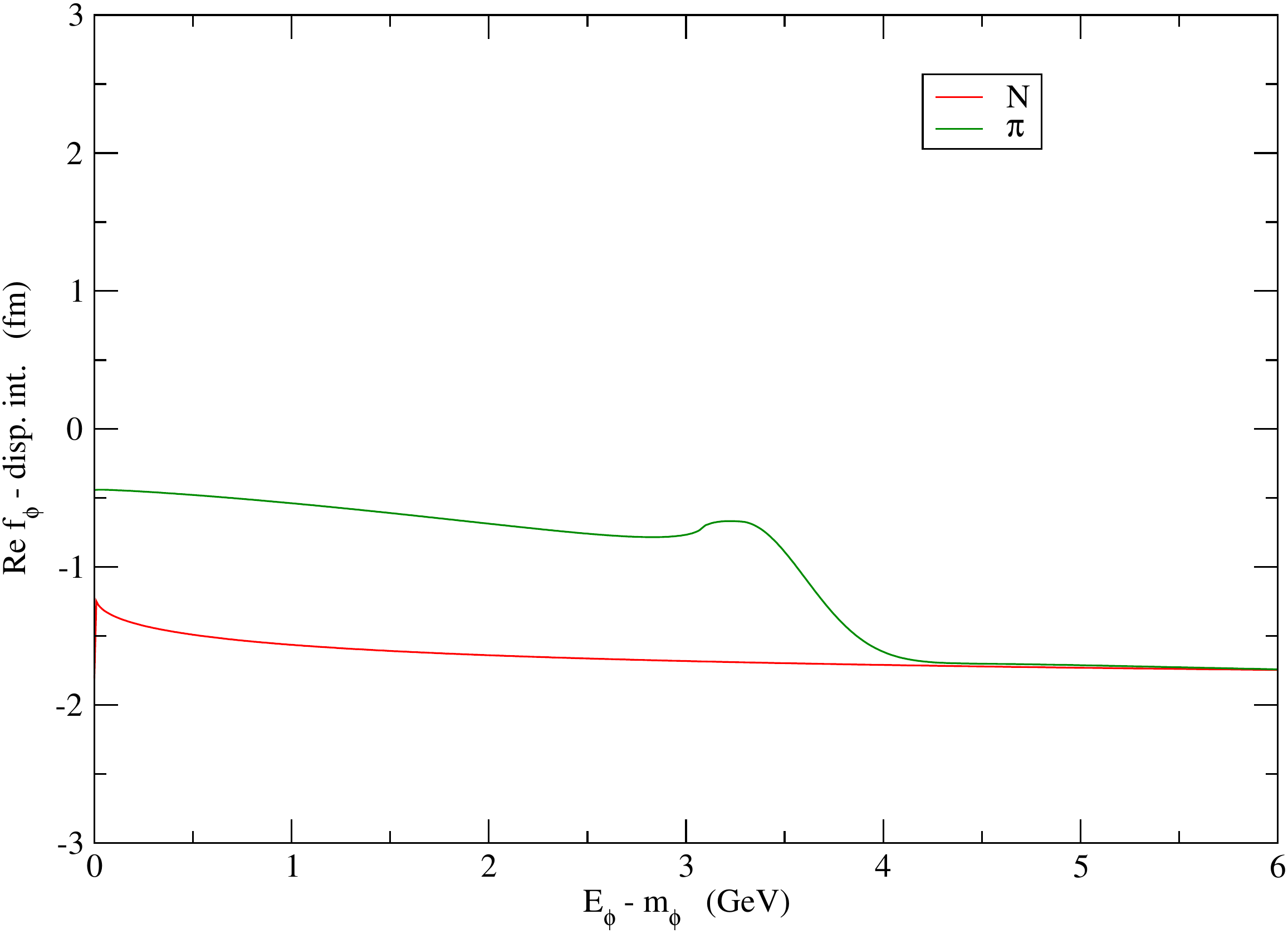}
\end{center}
\caption{Difference between the real part of the FSA in Fig. \ref{pic:re_im_f} (b) and the one calculated from the imaginary part of Fig. \ref{pic:re_im_f} (a) via the dispersion relation.}\label{pic:dispersion}
\end{figure}

After matching, one obtains the real an imaginary parts of the FSA as shown in Fig. \ref{pic:re_im_f}.
The kink in the imaginary part of the $\phi \pi$ FSA in Fig.\ref{pic:re_im_f} (a) is due to the $\rho(1450)$ resonance. The $\phi N$ curve is smooth, as expected, since the resonances contributing to the FSA have a small branching ratio. The real part of the $\phi \pi$ FSA has a change in sign as expected for a Breit-Wigner profile. Such a feature is not present in the $\phi N$ FSA, as the resonance contribution to the FSA is small. Fig. \ref{pic:dispersion} gives the plot of the real part minus the dispersion relation. 
The matching for the $\phi N$ FSA is better than the one for the $\phi \pi$ FSA as illustrated in Fig. \ref{pic:dispersion}. To explain this, note that the biggest deviation from a constant occurs right where low and the high energy FSA are matched. The $\phi \pi$ FSA is matched onto the Regge part slightly beyond $E_\phi \sim 4$ GeV. However, the $\phi N$ FSA is dominated by the Regge curve already below $E_\phi \sim 1$ GeV. The discrepancy from a constant is the most visible in the case of $\phi \pi$ as the matching occurs right around $E_\phi = 4$ GeV. The $\phi N$ case however shows quick convergence to a constant with only a slight variation at energies below 1 GeV. Of course, it should not be expected that our phenomenological approximations exactly obey the constraints that follow from the analytic properties of the FSA \cite{eletsky-belkacem-kapusta} and therefore, some deviations are expected.         

The dispersion relation is determined from the poles of the propagator with the on-shell self-energy, i.e. $M=m_\phi$. Hence the energy of $\phi$ meson takes the form:
\begin{equation}
E^2 = p^2 + m_\phi^2 + \Pi^{\rm net}_\phi(p)
\end{equation}
Since the self-energy is complex, one can decompose $E\left(p\right)$ as $E\left(p\right) = E_R\left(p\right) - i\Gamma(p)/2$.  The width is simply
\begin{equation}
\Gamma\left(p\right) = - \frac{{\rm Im} \left[\Pi^{\rm net}_\phi \left( p \right)\right] }{E_R\left( p \right)}\ ,
\end{equation} 
where 
\begin{equation}
2E_{R}\left( p \right) = m^2_\phi + p^2 + {\rm Re} \Pi^{\rm net}_\phi \left( p \right) + \sqrt{\left\{ m^2_\phi + p^2 + {\rm Re} \left[ \Pi^{\rm net}_\phi \left( p \right) \right]\right\}^2 + \left\{ {\rm Im} \left[\Pi^{\rm net}_\phi \left( p \right)\right]\right\}^2}
\end{equation}
The mass shift of the $\phi$ meson due to its interaction with the medium is
\begin{equation}
\Delta m_\phi \left( p \right) = \sqrt{m^2_\phi+ {\rm Re} \left[\Pi^{\rm net}_\phi \left( p \right)  \right]}- m_\phi
\end{equation}
We will consider nucleon densities of $n_N=0, \frac{1}{2}, 1$ and $2$ in units of equilibrium nuclear matter density ($n_0=0.16$ $\mathrm{nucleons}/\mathrm{fm}^3$). Table \ref{table:chemical_potentials} gives the nucleon chemical potentials at the temperatures and nucleon densities used in this work. Any contribution to the self-energy arising from antinucleons is ignored here.
\begin{table}[!ht]
\caption{Nucleon chemical potentials corresponding to different densities (in units of $n_0=0.16\ \mathrm{nucleons}/\mathrm{fm}^3$) and temperatures.} 
\centering 
\begin{tabular}{c | c | c | c } 
\hline 
Chemical potentials & $n=\frac{1}{2}n_0$ & $n=n_0$ & $n=2n_0$ \\
\hline\hline  
T=100 MeV & 675 MeV & 747 MeV & 821 MeV  \\
T=150 MeV & 437 MeV & 543 MeV & 650 MeV \\
[1ex] 
\hline 
\end{tabular}
\label{table:chemical_potentials} 
\end{table}

\begin{figure}[!ht]
\begin{center}
\includegraphics[scale=0.45]{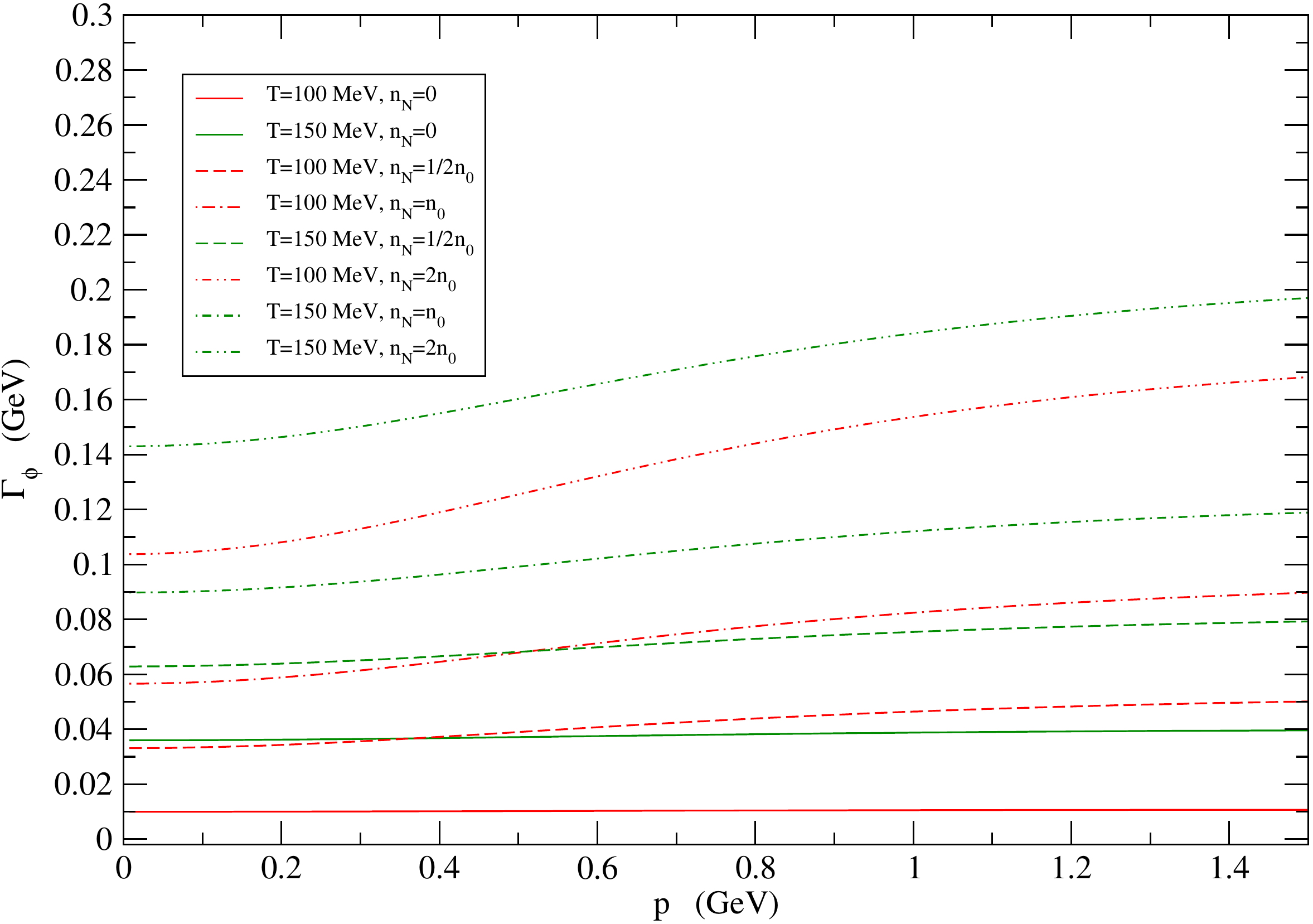}
\end{center}
\caption{Width of the $\phi$ meson as a function of $p$, $n_N$ and $T$. The results include four nucleon densities (namely $0$, $\frac{1}{2}n_0$, $n_0$, $2n_0$ with $n_0=0.16$$\mathrm{nucleons}/\mathrm{fm}^3$) and two temperatures (i.e. $T=100$ MeV, $T=150$ MeV).}\label{pic:Gamma}
\end{figure}

The plots giving the widths $\Gamma (p)$ are presented in Fig. \ref{pic:Gamma} for two temperatures and four nucleon densities. The widths are defined relative to the rest frame of the thermal system. 
Fig. \ref{pic:Gamma} shows that at vanishing nucleon densities and for a temperature of 100 MeV, the in-medium width of the $\phi$ meson is shifted very slightly from its vacuum value. At $T=150$ MeV, the width generated by collisions with pions is  about 40 MeV. Clearly, the behavior of $\Gamma(p)$ at zero nucleon density is insensitive to changes in momentum $p$. In contrast to the zero nucleon density results, at non-vanishing nucleon density we see that $\Gamma(p)$ develops a momentum dependence that increases with both increasing temperature and nucleon density. The in-medium width of the $\phi$ meson is quite different than the vacuum width and the magnitude of the in-medium width approaches that of the $\omega$ meson, but is not as pronounced as for the $\rho$ meson as can be seen in \cite{eletsky-belkacem-kapusta,martell-ellis}. 

\begin{figure}[!ht]
\begin{center}
\includegraphics[scale=0.45]{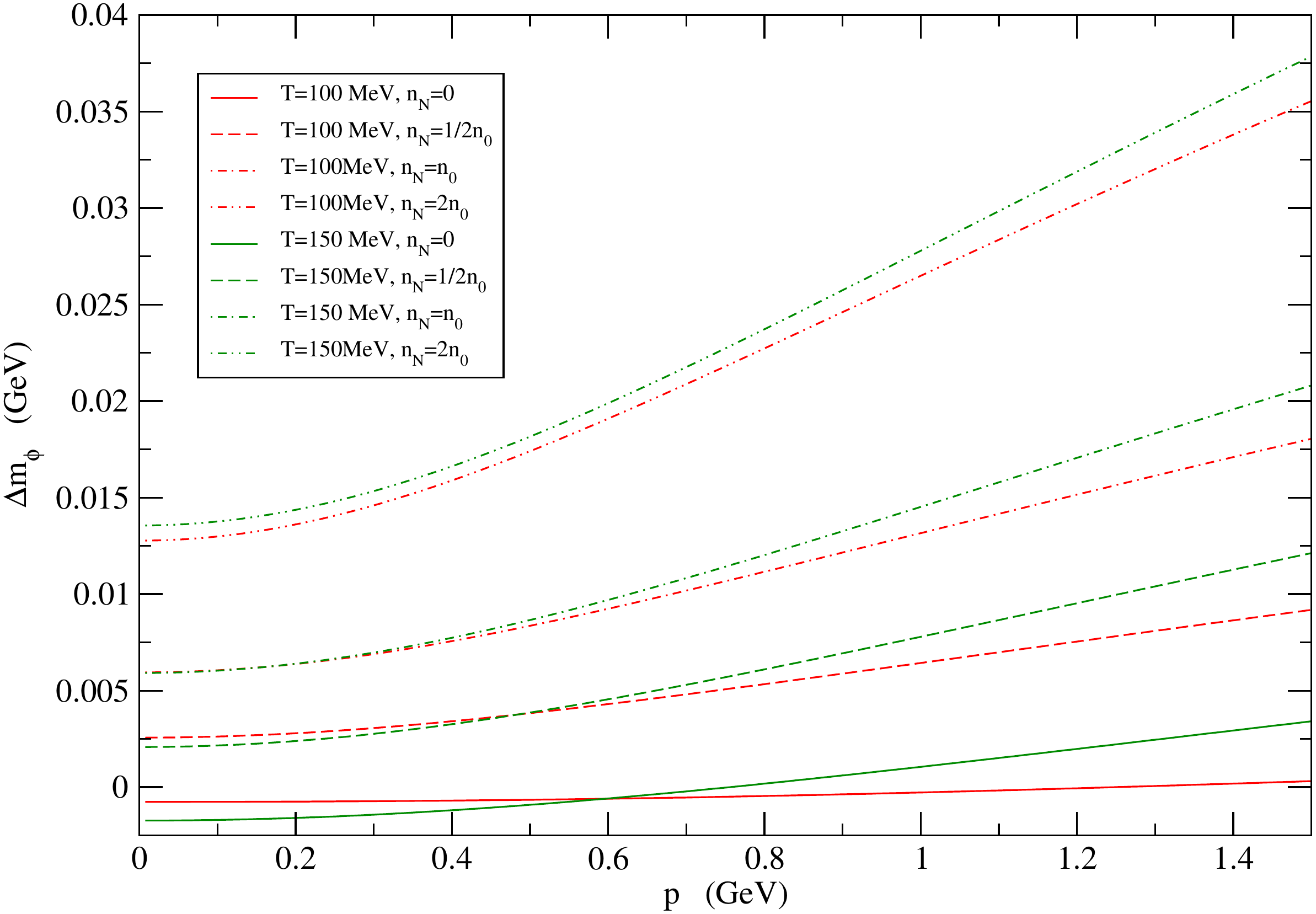}
\end{center}
\caption{Change in mass of the $\phi$ meson as a function of $p$, $n_N$ and $T$. The results include four nucleon densities (namely $0$, $\frac{1}{2}n_0$, $n_0$, $2n_0$ with $n_0=0.16$ $\mathrm{nucleons}/\mathrm{fm}^3$) and two temperatures (i.e. $T=100$ MeV, $T=150$ MeV).}\label{pic:Delta_m}
\end{figure}

The mass shift for the $\phi$ meson is presented in Fig. \ref{pic:Delta_m}. At p=0 GeV, $\Delta m$ is contained within the range of approximately -2.5 MeV to 15 MeV. This range greatly increases when $p=1.5$ GeV, and for $n_N=2$ and $T=150$ MeV gives a $\Delta m = 38$ GeV. Therefore, the change in mass overall is small, but an important message from this figure is that the change in mass is more density- than temperature-driven. Furthermore, interactions with pions only ($n_N=0$) give a small and mostly negative $\Delta m$, as was noted  previously in \cite{eletsky-belkacem-kapusta,martell-ellis}, while nucleons contribute to increasing the pole mass of the $\phi$. 
VMD \cite{gounaris-sakurai,gale-kapusta} allows us to relate the imaginary part of the $\phi$ meson propagator to the imaginary part of the photon  self-energy \cite{mclerran-toimela,weldon}, and then to the dilepton production rate, $E_{+} E_{-} dR/ d^3p_{+} d^3p_{-}$ \cite{ftft-book}. We shall let $p = 0.3$ GeV when evaluating $\Pi^{\rm net}_\phi$ so that our results can easily be compared with \cite{eletsky-belkacem-kapusta,martell-ellis}. 

\begin{figure}[!ht]
\begin{center}
\includegraphics[scale=0.45]{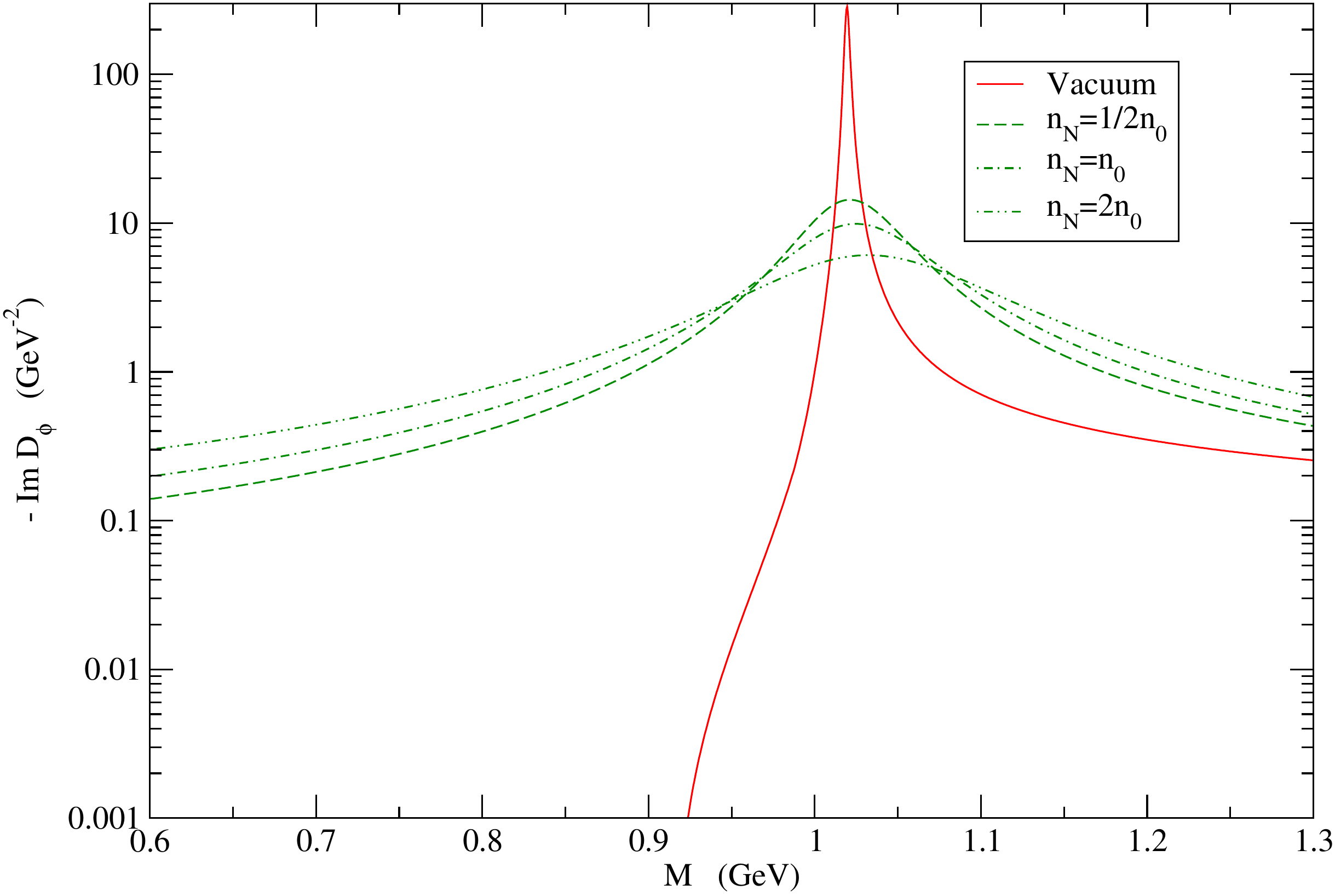}
\end{center}
\caption{The imaginary part of the $\phi$ meson propagator as a function of invariant mass $M$, for a momentum of $p$=0.3 GeV/c and a temperature of 150 MeV. The results include three nucleon densities (namely $\frac{1}{2}n_0$, $n_0$ and $2n_0$ with $n_0=0.16$ $\mathrm{nucleons}/\mathrm{fm}^3$). The vacuum imaginary part of the propagator is also included.}\label{pic:im_prop}
\end{figure}

The imaginary part of the propagator, directly proportional to the spectral density, is plotted as a function of $M$ in Fig. \ref{pic:im_prop} at $T=150$ MeV. Of course, at this point the vacuum part of the self-energy becomes $M$ dependent. 
One notices that the position of the peak in Fig. \ref{pic:im_prop} has not moved significantly whereas the width has been greatly enhanced. The width of the $\phi$ meson in medium is defined as 
\begin{equation}
\Gamma_{\phi}= -\frac{1}{m_{\phi}}{\rm Im}\left[ \Pi^{\rm net}_{\phi} (M^2=m^2_{\phi},p=0.3 \mathrm{GeV/c})\right]
\end{equation}
Our results are summarized in Table \ref{table:width_phi_t_mu}. 
\begin{table}[!ht]
\caption{The width of the $\phi$ meson at a temperature of 150 MeV and three different densities.} 
\centering 
\begin{tabular}{c c} 
\hline \hline 
Nucleon density &  Width \\
($n_0=0.16 \mathrm{nucleons}/\mathrm{fm}^3$) & (MeV)          \\ [0.5ex] 
\hline 
$\frac{1}{2}n_0$ & 68  \\
$n_0$ & 98  \\
$2n_0$ & 159 \\
[1ex] 
\hline 
\end{tabular}
\label{table:width_phi_t_mu} 
\end{table}
At $n_N=\frac{1}{2}n_0$ the width of the  of $\sim$68 MeV which is about a factor of 2 greater than the value obtained by Rapp \cite{rapp} at a higher temperature of 180 MeV and baryon density of half of the normal nuclear density. For $n_N=n_0$ the width is 98 MeV. This width is in agreement with the recent result published by van Hees and Rapp \cite{vhr} which have calculated the average width of the $\phi$ meson over their fireball evolution model to be $\sim80$ MeV. 

Holt and Haglin \cite{holt-haglin} have calculated the imaginary part of the propagator at zero nucleon density and temperatures of 170 MeV and 200 MeV. Our results seem to agree with Holt and Haglin \cite{holt-haglin} at a temperature $T=170$ MeV. Indeed, they have obtained a full width at half maximum (FWHM) of $\sim$50 MeV whereas our calculation gives a width of $\sim$58 MeV. The overall shape of the spectral density curves also seems to agree at that temperature. However, at a higher temperature of $T=200$ MeV, their prediction for the FWHM is about 300 MeV whereas we calculate a width around 100 MeV. Note that one expects that the hadronic degrees of freedom, as incorporated in both approaches, are anyways inappropriate at temperatures beyond the deconfinement phase transition.        

\begin{figure}[!ht]
\begin{center}
\includegraphics[scale=0.45]{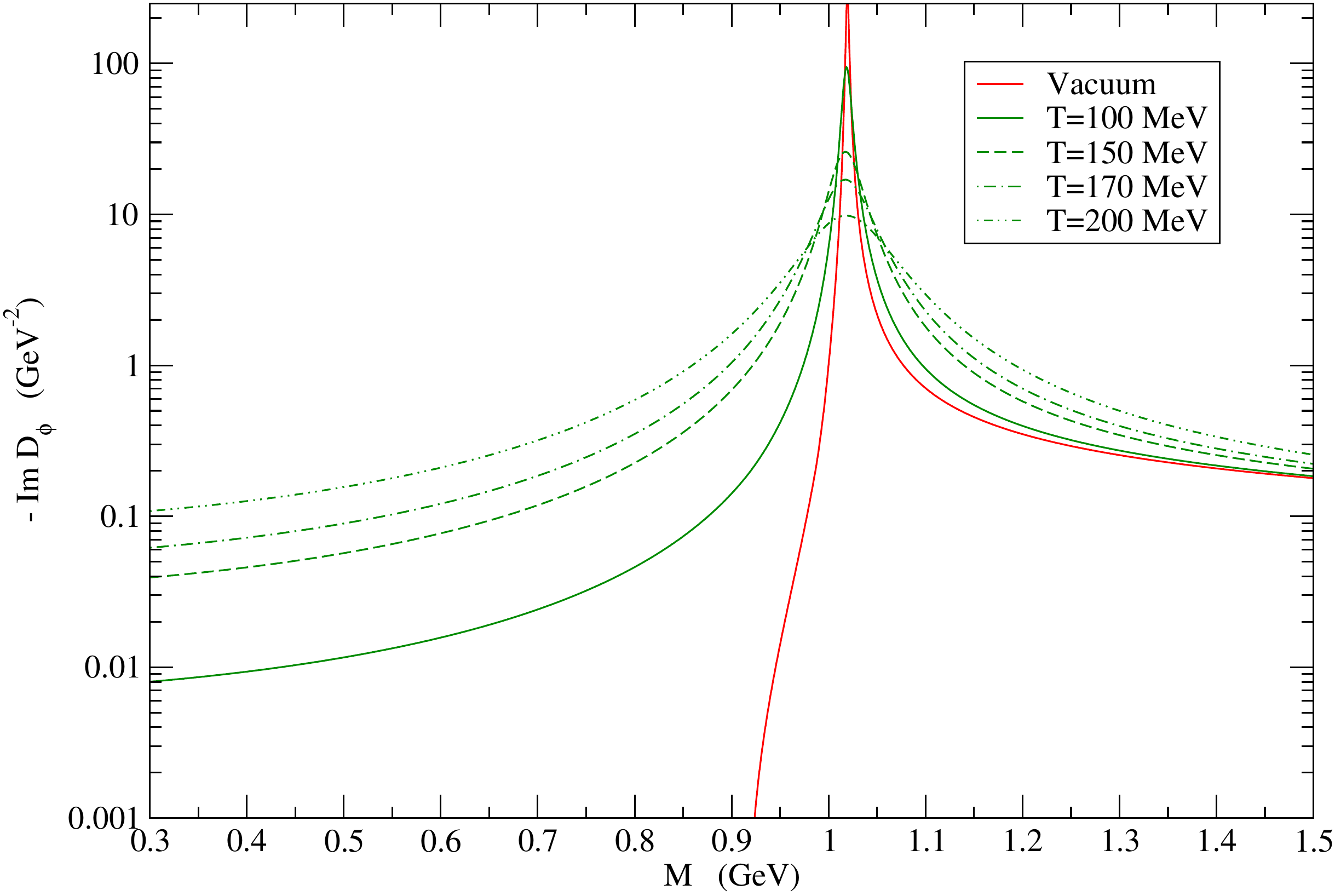}
\end{center}
\caption{The imaginary part of the $\phi$ meson propagator as a function of invariant mass $M$ for a momentum of 0.3 GeV/c and zero nucleon density. We present the vacuum result along with four temperatures (i.e. 100 MeV, 150 MeV, 170 MeV, and 200 MeV).}\label{pic:im_prop_haglin}
\end{figure}

In Fig. \ref{pic:im_prop_haglin}, we have displayed the imaginary part of the propagator at different temperatures and vanishing nucleon density; table \ref{table:width_phi_t} gives the corresponding widths. By comparing with Table \ref{table:width_phi_t_mu}, we observe once more that changes in width of the $\phi$ are influenced more by a density increase than a temperature change. This finding is in line with the conclusions of Refs. \cite{RW,vhr}.
\begin{table}[!ht]
\caption{The width of the $\phi$ meson at finite temperature and vanishing nucleon density.} 
\centering 
\begin{tabular}{c c} 
\hline \hline 
Temperature &  Width \\
(MeV) & (MeV)          \\ [0.5ex] 
\hline 
{\bf 0 }  & {\bf 4 }\\
100 & 11 \\
150 & 38 \\
170 & 58  \\
200 & 100  \\
[1ex] 
\hline 
\end{tabular}
\label{table:width_phi_t} 
\end{table}

\section{Conclusion}

We have calculated the mass shift and width broadening as a function of momentum for the $\phi$ meson using the scheme developed in \cite{eletsky-belkacem-kapusta,martell-ellis}. This approach has the great advantage to be model-independent since experimental data are taken as input to calculate the FSA. Our results show that pion interactions contribute to reduce the pole mass of the $\phi$ whereas nucleons act in the opposite way hence increasing the pole mass. The mass shift overall is small and about a few tens of MeV at most. Also, we have found a significant increase of the width of the $\phi$ meson from its vacuum value of 4.26 MeV. Hence the spectral density, which is related to the dilepton production, was considerably broadened. These  results will be  important for understanding possible in-medium modifications of the $\phi$ meson in strongly interacting environments. 

\begin{acknowledgments}

It is a pleasure to thank  J. I. Kapusta, P. Lichard, and A.T. Martell for usefull suggestions and discussions.  This work was supported in part by the  Natural Sciences and Engineering Research Council of Canada, and in part by the Fonds Nature et Technologies of Quebec.

\end{acknowledgments}


\begin{thebibliography}{00}
\bibitem{footnote} Actually, there is only one resonance has been observed to decays into a $\phi$ and a nucleon, the $\Lambda_{c}^{+}$. However, its partial width in this channel is small and has been neglected.
\end{thebibliography}
\end{document}